\documentclass[12pt]{iopart}

\usepackage{epsfig}

\begin{document}

\title[Stable oscillations of a predator-prey probabilistic cellular
automaton]
{Stable oscillations of a predator-prey probabilistic cellular
automaton: a mean-field approach.}

\author{T\^{a}nia Tom\'{e} and Kelly C de Carvalho}

\address{Instituto de F\'{\i}sica, 
Universidade de S\~{a}o Paulo \\
Caixa Postal 66318 \\
05315-970 S\~{a}o Paulo, S\~ao Paulo, Brazil}

\ead{ttome@if.usp.br}

\begin{abstract}

We analyze a probabilistic cellular automaton describing the
dynamics of coexistence of a predator-prey system. The individuals 
of each species are localized over the sites of a lattice and
the local stochastic updating rules are inspired on the processes of the 
Lotka-Volterra model. Two levels of mean-field approximations
are set up. The simple approximation is equivalent to an extended patch
model, a simple metapopulation model with patches colonized by
prey, patches colonized by predators and empty patches.
This approximation is capable of describing the limited 
available space for species occupancy.
The pair approximation is moreover able to describe two types of 
coexistence of prey and predators: one where population densities are
constant in time and another displaying self-sustained
time-oscillations of the population densities.
The oscillations are associated with limit cycles and arise
through a Hopf bifurcation. They are stable against changes in 
the initial conditions and, in this sense, they differ from the
Lotka-Volterra cycles which depend on initial conditions.
In this respect, the present model is biologically
more realistic than the Lotka-Volterra model.

\end{abstract}

\pacs{87.23.Cc, 05.65.+b, 05.70.Ln, 02.50.Ga}

\maketitle

\section{Introduction}

The simplest model exhibiting time-oscillations in a two-component
system is the model proposed independently by Lotka 
\cite{lotka1,lotka2,lotka3} 
and by Volterra \cite{volterra}. In this model the individuals of two
species are dispersed over an assumed homogeneous space. It
is implicitly assumed in this approach that any individual can interact with
any other one with equal intensity implying that
their positions are not taken into account.
The time evolution of the densities
of the two species in the Lotka-Volterra model
is given by a set of two ordinary differential equations 
\cite{haken,renshaw,hastings,ecology} and is set up in analogy with
the laws of mass-action.
Depending on the level of description wanted, 
the approach based on mass-action laws, contained
on the Lotka-Volterra model, suffices.
However, there are situations in which the coexistence 
takes place in a spatially heterogeneous habitat such that the  
population densities can be  very low in some regions. 
In this case we need to proceed beyond the
mass-law equations and consider the space structure of the
habitat. In other words, it becomes necessary 
to analyze the coexistence by taking explicitly into account 
spatial structured models. 

In fact, the role of space in
the description of population biology problems  has
been recognized by several authors in the last years 
\cite{tainaka,sat,durrett,hanski,sat2,tilman,fracheburg,durrett2,
lipowsky,ovaskanien,aguiar,kelly,tainaka1,szabo,
stauffer,kelly1,mobilia,arashiro}.
In a very clear manner, Durrett and Levin \cite{durrett}
have pointed out that the modelling of population dynamics systems
which are spatially distributed
by interacting particle systems \cite{durrett,liggett,marro,ttmjo}
is the appropriate theoretical approach that is able
to give the more complete description of the problem.
We include in this approach probabilistic cellular 
automata (PCA) \cite{ttmjo,tome,tome4}, which will concern us here. 
We refer to interacting particle systems
and PCA as stochastic lattice models. They are
both Markovian processes defined by discrete stochastic
variables residing on the sites of a lattice;
the former being a continuous time process and the latter a
discrete time process.

In the present work we study the coexistence and the emergence of stable
self-sustained oscillations in a predator-prey system by considering
a PCA previously studied by numerical simulations \cite{kelly1,arashiro}. 
This PCA is defined by local rules, similar to
the ones of the contact process \cite{liggett}, that are capable 
of describing the interaction between prey and predator. 
Here, we focus on the analysis of the PCA by
means of dynamic mean-field approximations
\cite{sat,marro,ttmjo,tome,dickman,tome1}. 
In this approach the equations for the time evolution
of correlations of various orders are truncated at a 
certain level and 
high order correlations of sites are written in terms of
small order correlations. The simplest approximation is
the one in which all correlations are written in terms of
one-site correlation, called simple approximation. 
In a more sophisticated approximation,
called pair approximation \cite{sat,ttmjo},
any correlation is written in terms of one-site and two-site
correlations. 

The simple mean-field approximation is capable of predicting
coexistence of individuals in a stationary state where the
densities of each species, and of empty sites, are constant.
However, it is not capable of predicting possible time 
oscillating behavior of the population densities and we have 
proceed to the next order of mean-field approximation.
The simple approximation, on the other hand, can be placed in an explicit
correspondence  with a patch model \cite{hastings,hanski,levins} where
unoccupied patches can be colonized by prey and patches 
occupied by prey can be colonized by predators that
in turn may become extinct. In this approximation the PCA can 
be seen as an extended version of the Lotka-Volterra model which
includes an extra logistic term related to the empty sites.

The pair-mean field approximation is 
able to predict possible time oscillating
behavior of the population densities
that are self-sustained and are
attained thorough Hopf bifurcations. 
This is in contrast with the Lotka-Volterra model which presents
no stable oscillations but exhibits instead infinite cycles 
that are associated to different initial conditions. 
However, from the biological point of view, one does not expect 
that a small variation in the initial densities of prey and predator result
in different amplitudes of oscillations. Within our approach,
a PCA treated in the pair-approximation, 
the oscillations are associated with limit cycles what
mean to say that they are stable against the changes in the initial 
conditions. According to our point of view, the pair-approximation,
in which the correlation between neighboring sites are treated exactly, 
provides a basic description of the predator-prey 
spatial interactions. For this reason, we will refer to the PCA in this
approximation as a quasi-spatial-structured model. 

\section{Model}

\subsection{Probabilistic cellular automaton}

We consider interacting particles living on the sites of
a lattice and evolving in time according to Markovian local rules. 
The lattice is the geometrical object that plays the
role of the spatial region occupied by particles, in a general
case, or by individuals of each species in the present case. 
The lattice sites are the possible locations for the individuals.
Each site can be either empty or occupied by one individual of 
different species and a stochastic
variable $\eta _{i}$ is introduced to describe the state of each site
at a given instant of time. The state of
the entire system is denoted by 
$\eta =(\eta_1,\dots,\eta _i,\ldots,\eta_N)$ where $N$ is
the total number of sites. 
The transition between the states is governed by the interactions
between neighbor sites in the lattice and by a synchronous dynamics. 

The probability $P^{(\ell)}(\eta)$ of configuration $\eta$ at 
time step $\ell$ evolves according to the Markov chain equation 
\begin{equation}
P^{(\ell+1)}(\eta)= \sum_{\eta} W(\eta|\eta^{\prime })
P^{(\ell)}(\eta ^{\prime }),  
\label{1}
\end{equation}
where the summation is over all the microscopic configurations 
of the system, and $W(\eta |\eta ^{\prime })$ is the
conditional transition probability from state $\eta^{\prime }$ at
time $\ell $ to state $\eta$ at time $\ell+1$. This transition
probability does not depend on time and contains all the information
about the dynamics of the system. 
Taking into account that all the sites are simultaneously updated,
which is the fundamental property of a PCA,
the transition probability can be factorized and written in 
the form \cite{ttmjo,tome}
\begin{equation}
W(\eta |\eta ^{\prime })
=\prod_{i=1}^{N}w_{i}(\eta _{i}|\eta ^{\prime }),
\label{2}
\end{equation}
where $w_{i}(\eta _{i}|\eta ^{\prime })$ is the conditional transition
probability that site $i$ takes the state $\eta _{i}$ given that the whole
system is in state $\eta ^{\prime }$. Being a probability
distribution, the quantity $w_{i}(\eta _{i}|\eta ^{\prime })$
must satisfy the following properties: 
$w_{i}(\eta _{i},\eta^{\prime })\geq 0$ and
\begin{equation}
\sum_{\eta _{i}}w_{i}(\eta _{i}|\eta ^{\prime })=1.
\end{equation}

The average of any state function $F(\eta )$ is evaluated by
\begin{equation}
\langle F(\eta )\rangle_{\ell }=\sum_{\eta }F(\eta )P^{(\ell)}(\eta ).  
\label{3}
\end{equation}
The time evolution equation for $\langle F(\eta )\rangle$ 
is obtained from definition (\ref{3}) and equation (\ref{1}). 
For example, we can derive the equations
for the time evolution of densities and two-site correlations.

\subsection{Predator-prey probabilistic cellular automaton}

To model a predator-prey system by a PCA,
the stochastic variable $\eta_i$ associated to site $i$ will 
represent the occupancy of the site by one prey, 
or the occupancy by one predator or the vacancy
(a site devoid of any individual). 
The variable $\eta_i$ is assumed to take the value 0, 1, or 2, according 
to whether the site is empty (V), occupied by a prey individual (H) 
or by a predator (P), respectively. That is,
\begin{equation}
\eta_i = \left\{
\begin{array}{ll}
0, & {\rm empty\,(V)}, \\
1, & {\rm prey\,(H)}, \\
2, & {\rm predator\, (P),}
\end{array}
\right.
\end{equation}
which defines a three-state per site PCA.

\begin{figure}
\centering
\epsfig{file=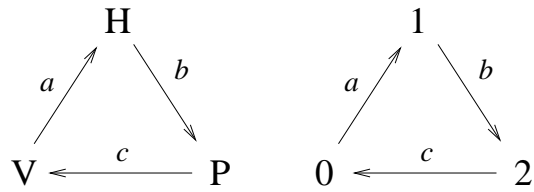,width=7cm}
\caption{Transitions of the predator-prey model. 
The three states are: prey or herbivorous
(H or 1), predator (P or 2) and empty or vegetable (V or 0).
 The allowed transitions obey the cyclic order shown.}
\label{vhp}
\end{figure}

The stochastic rules, embodied in the transition
rate $w_i(\eta_i|\eta^\prime)$, are set up according to the following 
assumptions. (a) The space is homogeneous, which means to say that
no region of the space will
be privileged against the others, that is, in principle the individuals have
the same conditions of surveillance in any space region. 
(b) The space is isotropic, which means to say that there is 
no preferential direction in this space for any interaction. 
(c) The allowed transitions between states are only the
ones that obey the cyclic order shown in figure \ref{vhp}. Prey can only
born in empty sites; prey can give place to a predator, in a process where
a prey individual dies and a predator is instantaneously born; 
finally a predator can
die leaving an empty site. The empty sites are places where prey can
proliferate and can be seen as the resource for prey surveillance.
The death of predators complete this cycle, reintegrating to the system the
resources for prey.

The predator-prey PCA
has three parameters: $a$, the probability of birth of prey,
$b$, the probability of birth of predator and death of prey, and $c$, the
probability of predator death. Two of the process are catalytic: the
occupancy of a site by prey or by a predator is conditioned, respectively,
to the existence of prey or predator in the neighborhood of the site. The
third reaction, where predator dies, is spontaneous, that is, it occurs,
with probability $c$, independently of the neighbors of the site. 
We assume that 
\begin{equation}
a+b+c=1,
\label{abc}
\end{equation}
with $0\leq a,b,c\leq 1$.

The transition probabilities of the predator-prey PCA
are described in what follows:

(a) If a site $i$ is empty, $\eta _{i}=0$, and there is at least one
prey in its first neighborhood there is a favorable condition for the birth
of a new prey. The probability of site $i$ being occupied
in next time step by a prey is proportional to the
parameter $a$ and to the number of prey that are in the first neighborhood
of the empty site.

(b) If a site is occupied by a prey, $\eta_i=1$,
and there is at least one
predator in its first neighborhood then the site has a probability of being
occupied by a new predator in the next instant of time. In this process the
prey dies instantaneously. The transition probability
is proportional to the parameter $b$ and the 
number of predators in first neighborhood of the site.

(c) If site $i$ is occupied by a predator, $\eta_i=2$,
it dies with probability $c$. 

The transition probabilities associated to the three processes 
above mentioned can be summarized as follows:
\begin{equation}
w_i(0|\eta)=
c \delta(\eta_i,2) 
+ [1-f_i(\eta)] \delta(\eta_i,0),  
\label{4}
\end{equation}
\smallskip
\begin{equation}
w_i(1|\eta)=f_i(\eta) \delta(\eta_{i},0) 
+ [1-g_i(\eta)]\delta(\eta_{i},1),  
\label{5}
\end{equation}
\smallskip
\begin{equation}
w_{i}(2|\eta)=g_i(\eta)\delta (\eta _{i},1) 
+(1-c)\delta (\eta _{i},2),  
\label{6}
\end{equation}
\smallskip
where
\begin{equation}
f_i(\eta) =
\frac{a}{4}\sum_k \delta(\eta_k,1),
\qquad\qquad
g_i(\eta) =
\frac{b}{4}\sum_k \delta(\eta_k,2),
\end{equation}
and the summation is over the four nearest neighbors of site $i$ in a
regular square lattice. The notation $\delta (x,y)$ stands for the
Kronecker delta function. These stochastic local rules, 
when inserted in equation (\ref{2}), define the dynamics of the 
PCA for a predator-prey system.

The present stochastic dynamics predicts the existence of states,
called absorbing states, in which the system becomes trapped.
Once the system has entered such a state it cannot escape from it
anymore remaining there forever. 
There are two absorbing states. One of them is the empty lattice.
Since the predator death is spontaneous, a
configuration where just predators are present is not
stationary. This situation happens whenever the prey have been extinct.
In this case the predator cannot reproduce anymore and also get extinct,
leaving the entire lattice with empty sites. 
The other absorbing
state is the lattice full of prey. This situation occurs if there are few
predators and they become extinct. The remaining prey will then
reproduce without predation filling up the whole lattice.
The existence of absorbing stationary states is an 
evidence of the irreversible character of the model or, in other
words, of the lack of detailed balance \cite{ttmjo}. 
However, the most interesting states, the ones that we
are concerned with in the present study, are the active states 
characterized by the coexistence of prey and predators. 

\subsection{Time evolution equations for state functions}

We start by defining the densities, which are the one-site correlations, 
and the two-site correlations. These quantities will be useful in our 
mean-field analysis to be developed below. 
The density of prey, predator, and empty sites at time step $\ell$ 
are defined thought the expressions
\begin{equation}
P_i^{(\ell)}(1)=\langle\delta (\eta _{i},1)\rangle_{\ell}, 
\end{equation}
\begin{equation}
P_i^{(\ell)}(2)=\langle\delta (\eta _{i},2)\rangle_{\ell}, 
\end{equation}
\begin{equation}
P_i^{(\ell)}(0)=\langle\delta (\eta _{i},0)\rangle_{\ell}.  
\end{equation}
The evolution equations for the above densities are obtained from
their definitions as state functions, as given by
equation (\ref{3}), and by using the evolution equation for 
$P^{(\ell)}(\eta)$, given by equation (\ref{1}). 
The resulting equations can be formally written as 
\begin{equation}
P_i^{(\ell+1)}(1) =\langle w_{i}(1|\eta)\rangle_{\ell },  
\end{equation}
\begin{equation}
P_i^{(\ell+1)}(2) = \langle w_{i}(2|\eta)\rangle_{\ell },
\end{equation}
\begin{equation}
P_i^{(\ell+1)}(0) = \langle w_{i}(0|\eta)\rangle_{\ell },
\end{equation}
where the transition probabilities for this model are given in equations 
(\ref{4}), (\ref{5}) and (\ref{6}). 

The correlation between a prey localized at site $i$ 
and a predator localized at site $j$ at time step $\ell$ is defined by
\begin{equation}
P_{ij}^{(\ell)}(1,2)=\langle\delta(\eta_i,1)\delta(\eta_j,2)
\rangle_{\ell}.  
\end{equation}
The other two-site correlations are defined similarly.
The time evolution equation for the correlation of two neighbor sites 
$i$ and $j$, one being occupied by a prey and the other 
by a predator, is given by
\begin{equation}
P_{ij}^{(\ell+1)}(1,2)=
\langle w_{i}(1|\eta)w_{j}(2|\eta)\rangle_{\ell}.
\label{8}
\end{equation}
The other two-site evolution equations are given by similar 
formal expressions. We
can also derive equations for three-site correlations. 
Since we are interested here on approximations in which 
only the one-site and two-site correlations should be
treated exactly, the above equations suffice. 

We call the attention to the fact that equation (\ref{8}) includes the product
of two transition probabilities. This is a consequence of the synchronous
update of the PCA which allows that both neighboring sites $i$ and $j$
have their states changed at same time step. 
This situation does not occur when we consider a continuous time
one-site dynamics. Therefore, although local interaction in the present
PCA and in the continuous time model considered in reference \cite{sat}
are the same, the predator-prey system evolves according to different
global dynamics which leads to different time evolution equations for the
densities and the correlations.

The exact evolution equations for the one-site correlations are
\begin{equation}
P_j^{\,\prime}(1) = \frac{a}{\zeta} \sum_i P_{ji}(01) - 
\frac{b}{\zeta} \sum_i P_{ji}(12) + P_j(1),
\end{equation}
\begin{equation}
P_j^{\,\prime}(2) = \frac{b}{\zeta} \sum_i P_{ji}(12) + (1-c) P_j(2),
\end{equation}
where the summation in $j$ is over the $\zeta$ nearest neighbors of site $i$.
To simplify notation we are using unprimed and primed quantities 
to refer to quantities taken at time $\ell$ and $\ell+1$,
respectively.

The exact evolution equations for the correlations of two nearest
neighbor sites $j$ and $k$ are
\begin{eqnarray}
P_{jk}^{\,\prime}(01) 
& = & 
\frac{a}{\zeta}\sum_{n(\neq j)} \left[ P_{jkn}(001) -\frac{a}{\zeta} 
\sum_{i(\neq k)} P_{ijkn}(1001) \right] 
\nonumber \\
& + & (1-\frac{a}{\zeta})\left[ P_{jk}(01) -\frac{b}{\zeta} 
\sum_{n(\neq j)} P_{jkn}(012) \right]
\nonumber \\
& - & \frac{a}{\zeta}\sum_{i(\neq k)} \left[ P_{ijk}(101) -\frac{b}{\zeta} 
\sum_{n(\neq j)} P_{ijkn}(1012) \right] 
\nonumber \\
& + & c\left[ (1-\frac{b}{\zeta}) P_{jk}(21) -\frac{b}{\zeta} 
\sum_{n(\neq j)} P_{jkn}(212) \right]
\nonumber \\
& + & \frac{a}{\zeta}\,c\sum_{n(\neq j)} P_{jkn}(201),
\end{eqnarray}
\begin{eqnarray}
P_{jk}^{\,\prime}(12) & = & 
\frac{ab}{\zeta^2}\sum_{n(\neq j)} \left[ P_{jkn}(012) + 
\sum_{i(\neq k)} P_{ijkn}(1012) \right] 
\nonumber \\
& + & \frac{b}{\zeta}\sum_{n(\neq j)} \left[ P_{jkn}(112) 
-\frac{b}{\zeta} \sum_{i(\neq k)} P_{ijkn}(2112) \right] 
\nonumber \\
& + & (1-c) \left[ (1-\frac{b}{\zeta}) P_{jk}(12)  
-\frac{b}{\zeta}\sum_{i(\neq k)} P_{ijk}(212) \right]
\nonumber \\
& + & \frac{a}{\zeta}\,(1-c)\sum_{i(\neq k)} P_{ijk}(102),
\end{eqnarray}
and
\begin{eqnarray}
P_{jk}^{\,\prime}(02) & = & 
\frac{b}{\zeta}\sum_{n(\neq j)} \left[ (1-\frac{a}{\zeta})P_{jkn}(012) 
-\frac{a}{\zeta} \sum_{i(\neq k)} P_{ijkn}(1012) \right] 
\nonumber \\
& + & (1-c) \left[ c P_{jk}(22) + P_{jk}(02) 
-\frac{a}{\zeta} \sum_{i(\neq k)} P_{ijk}(102) \right] 
\nonumber \\
& + & \frac{b}{\zeta}\,c \left[ P_{jk}(21) 
+ \sum_{n(\neq j)} P_{jkn}(212) \right],
\end{eqnarray}
where the summation in $i$ is over the nearest neighbors of $j$ 
and the summation in $n$ is over the nearest neighbors of $k$.

\section{Mean-field approximation}

\subsection{One and two site approximations}

The evolution equation for a density in any interacting particle system
which evolves in time according to local interaction rules always contains
terms related to the correlations between neighbor sites in a lattice. The
evolution equations for the correlations of two neighbor sites includes the
correlation of clusters of three or more sites in the lattice and so on. In
this way we can have an infinite set of coupled equations for the
correlations which is equivalent to the evolution equation for the
probability $P^{(\ell)}(\eta )$, described in equation (\ref{1}) for the
automaton. The scope of the dynamic mean-field approximation consists in the
truncation of this infinite set of coupled equations 
\cite{tome,tome4,dickman,tome1}.

\begin{figure}
\centering
\epsfig{file=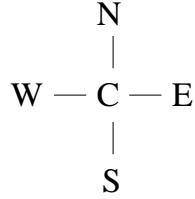,width=2.5cm}
\caption{A site (C) of the square lattice and its four nearest
neighbor sites (N, E, W, S).}
\label{news}	
\end{figure}

The lowest order dynamic mean-field approximation is the one where the
probability of a given cluster is written as the product of the
probabilities of each site. That is, all the correlations between sites in
the cluster are neglected. For example, let us consider the cluster
constituted by a center (C) site and its first
neighboring sites to the north (N), south (S), 
east (E) and west (W) as shown in figure \ref{news}.

Within the one-site approximation the probability $P(N,E,W,S,C)$ 
corresponding to the cluster shown in figure \ref{news} is approximated by
\begin{equation}
P(N,E,W,S,C) = P(N)P(E)P(W)P(S)P(C),  
\label{smf}
\end{equation}
where $P(X)$, $X=N,E,W,S,C$ 
are the one-site probabilities corresponding to each site.
For some stochastic dynamics models this approximation is able 
to give qualitative results that are in agreement with the expected results. 

In order to get a better approximation we must include fluctuations.
The simplest mean-field approximation that includes correlations 
is the pair-mean field approximation. This approximation is better
explained by taking again, as an example,
the cluster constituted by a center site which
and its four nearest neighbors, shown above.
Within the pair-approximation the conditional probability 
$P(N,E,W,S\,|\,C)$ is approximated by 
\begin{equation}
P(N,E,W,S\,|\,C) = P(N\,|\,C)P(E,|\,C)P(W\,|\,C)P(S\,|\,C),
\end{equation}
that is, the conditional probability $P(N,E,W,S\,|\,C)$  
is written in terms of
the product of the conditional probabilities $P(X|C)$,
$X=N,E,W,S$. Now using the
definition of conditional probability we have
\begin{equation}
\frac{P(N,E,W,S,C)}{P(C)} = \frac{P(N,C)}{P(C)}\frac{P(E,C)}{P(C)}
\frac{P(W,C)}{P(C)}\frac{P(S,C)}{P(C)},
\label{aproxpares}
\end{equation}
or
\begin{equation}
P(N,E,W,S,C) = \frac{P(N,C)P(E,C)P(W,C)P(S,C)}{[P(C)]^3}.  
\label{pairmf}
\end{equation}
We see that the resulting probability is
written as a function of two-site correlations $P(X,C)$, and
the one-site correlation $P(C)$.

\subsection{Patch model}

The simple mean-field approximation of the predator-prey PCA
describes exactly the same properties of an extended Levins 
patch model \cite{hastings,levins}. That is,
the PCA with local rules similar to
the contact process becomes, in the simple mean-field approximation,
analogous to the Levins model for metapopulation with empty patches, patches
colonized by prey and patches colonized by predators.

In the one-site mean-field approximation
we consider that the probability of
any cluster of sites can be written as the product of the probabilities of
each site, as in equation (\ref{smf}). Using this approach, 
and writing $x=P_i(1)$, $y=P_i(2)$, and
$z=P_i(0)$
it can be seen
that the set of equations can be reduced to the following
two-dimensional map \cite{arashiro}
\begin{equation}
x^\prime=x + a x z -bx y,  
\label{9}
\end{equation}
which is an evolution equation for prey density $x$, and 
\begin{equation}
y^\prime=y + bx y-cy,  
\label{10}
\end{equation}
which is an evolution equation for predator density $y_\ell$.
Notice that
\begin{equation}
z=1-x-y.
\end{equation}

The fixed point of this map are those that
represent the stationary solutions
$x^\prime=x$ and $y^\prime=y$,
and they correspond to the three following solutions
$x_1=0$, $y_1=0$, and $x_2=1$, $y_2=0$,  
and
$x_3 = a/b$, $y_3= (1-c/b)/(1+b/a)$.  
The first solution corresponds to an absorbing states where
both species have been extinct. The second corresponds to
an absorbing state where predators have extinct. The third solution
corresponds to an active state where prey and predator coexist.

Due to the constraint (\ref{abc}), the parameters $a$, $b$ and $c$
are not all independent and only two can be chosen as independent. For this
reason it is convenient to introduce 
the following parametrization \cite{sat}
\begin{equation}
a=\frac{1-c}{2}-p,
\qquad\qquad
b=\frac{1-c}{2}+p, 
\label{parameters}
\end{equation}
and consider $p$ and $c$ as the independent variables. 
The parameter $p$ is such that $-1/2\leq p\leq 1/2$ and 
$0\leq c\leq 1$ as before.
This parametrization will useful in the determination of 
the different phases displayed by the model.

\begin{figure}
\centering
\epsfig{file=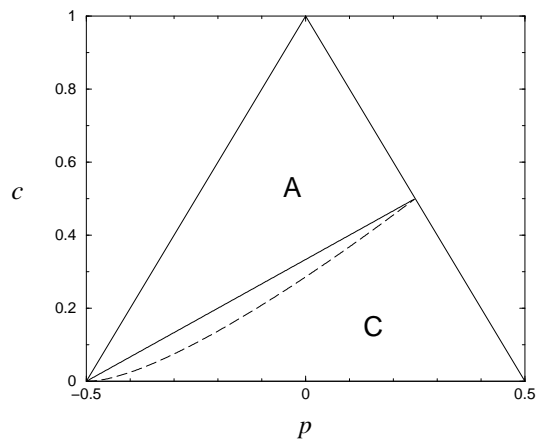,width=7cm}
\caption{Phase diagram of the patch model. The continuous line
represents the transition, $c_1(p)$,
between the prey absorbing (A) state
and the active species coexistence (C) state. The dashed line
separates the two asymptotic time behavior of the active state.}
\label{diag-one}	
\end{figure}

A linear stability analysis reveals that solution the 
$(x_{1},y_{1})$ is a hyperbolic saddle point for any set of 
the parameters $a,b$ and $c$ and so it is always unstable. 
The empty absorbing state will never be reached.
A linear stability analysis also shows that 
the solution $(x_{2},y_{2})$ is a stable node in the following
region of the phase diagram $c > c_1$ where 
\begin{equation}
c_1(p) = \frac{1}{3}(1+2p).
\end{equation}
The active solution is stable in the region $c < c_1$
and is attained in two ways: by an asymptotic stable
focus, where the successive interactions of the map show damped
oscillations; or trough an asymptotic stable node. 
In the phase diagram of figure \ref{diag-one} we show 
the transition line between the prey absorbing state and
the active state given by $c=c_1$.

\begin{figure}
\centering
\epsfig{file=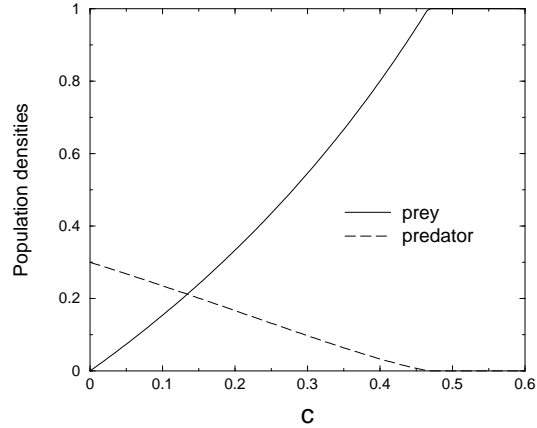,width=7cm}
\caption{Densities of predator and prey as functions
of the parameter $c$ for $p=0.2$, for the patch model.}
\label{dens-one}
\end{figure}

In figure \ref{dens-one} it is shown the behavior of the densities against the
parameter $c$, the probability of predators death, for the special case 
$p=0.2$. In terms of phase transitions what
happens is that in the phase diagram there is a transition line
separating the absorbing prey phase and the active phase which is
characterized by constant and nonzero densities of prey and predator.

We may conclude that the mean-field approximation for the
predator-prey probabilistic cellular automaton with rules (\ref{4}),
(\ref{5}), and (\ref{6}) 
is capable to show, under a robust set of control parameters, that prey
and predators can coexist without extinction. However the map 
defined by equations (\ref{9}) and (\ref{10}) is not able
to describe self-sustained oscillations of species population densities. 

\subsection{Quasi-spatial model}

In order to find if oscillations in the species populations can be described
within a mean-field approach we consider a more sophisticated approximation,
the pair-approximation, where correlations of two neighbor sites are
included in the time evolution equations for the densities. This is the
lowest order mean-field approximation which takes into account the spatial
localization of neighboring individuals.

In this analysis we will maintain the correlations of one site and the
correlations of two-sites in the equations. Correlations of three
and four neighbor sites will be approximated by means of equation
(\ref{pairmf}).
With these approximation the model is described by the
following set of five coupled equations
\begin{equation}
x^\prime=au-bv+x,
\label{pair1}
\end{equation}
\begin{equation}
y^\prime=bv+(1-c)y,
\label{pair2}
\end{equation}
\begin{eqnarray}
u^\prime
&=&\alpha a [\frac{q u}{z}-\alpha a \frac{q u ^{2}}{z^{2}}]+
[(1-\beta a)-\alpha a \frac{u}{z}] [u-\alpha b\frac{u v}{x}]
\nonumber \\
&+&\alpha ac\frac{w u}{z}
+c[(1-\beta b)v-\alpha b \frac{v^2}{x}],
\label{pair3}
\end{eqnarray}
\begin{eqnarray}
v^\prime
&=&\alpha b[\beta a \frac{u v}{x} +\alpha a\frac{u^2 v}{z x}]
+\alpha a (1-c) \frac{w u}{z}
\nonumber \\
&+&\alpha b [\frac{r v}{x}-\alpha b\frac{r v^2}{x^2}]
+(1-c)[(1-\beta b)v-\alpha b \frac{v^2}{x}],
\label{pair4}
\end{eqnarray}
and
\begin{eqnarray}
w^\prime
&=&\alpha b[(1-\beta a)\frac{u v}{x} -\alpha a \frac{u^2 v}{z x}]
+(1-c)[w-\alpha a \frac{u w}{z}]
\nonumber \\
&+&c[\beta b v+\alpha b \frac{v^2}{x}]+c(1-c)s,  
\label{pair5}
\end{eqnarray}
where $\alpha$ and $\beta$ are numerical fractions defined by
$\alpha =(\zeta -1)/\zeta$ and $\beta=1/\zeta$ where $\zeta$
is the coordination number of the lattice.
For the present case of a square lattice, $\zeta=4$ so that
$\alpha=3/4$ and $\beta=1/4$. 
We are using the following notation:
$u=P(0,1)$, $v=P(1,2)$, and $w=P(0,2)$ 
and also $r = P(1,1)$, $q = P(0,0)$
and $s = P(2,2)$.
The last three correlations are not independent but 
are related to others by
\begin{equation}
r = x - u - v,
\end{equation}
\begin{equation}
q = z - u - w,
\end{equation}
\begin{equation}
s = y - v - w.  
\label{14a}
\end{equation}
We used the properties $P(1,0)=P(0,1)$,
$P(1,2)=P(2,1)$ and $P(2,0)=P(0,2)$,
that follows from the assumption that
space is isotropic and homogeneous.

\begin{figure}
\centering
\epsfig{file=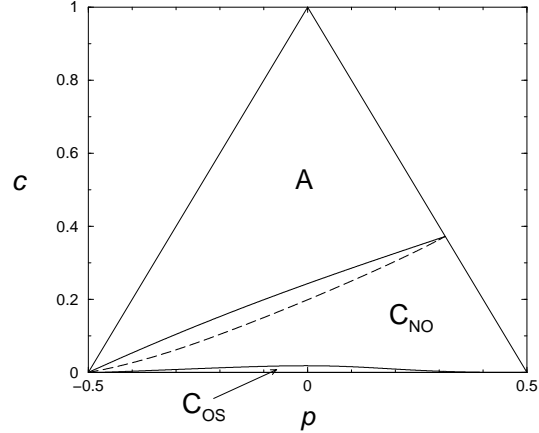,width=7cm}
\caption{Phase diagram of the quasi-spatial model.
The upper continuous line represents the transition, $c_1(p)$,
between the prey absorbing (A) state
and the nonoscillating coexistence (C$_{\rm NO}$) state. The lower
continuous line represents the transition, $c_2(p)$,
between the 
nonoscillating coexistence and the oscillating (C$_{\rm OS}$)
coexistence state. 
The dashed line separates the two asymptotic time behavior of 
the nonoscillating coexistence state.}
\label{diag-two}	
\end{figure}

\begin{figure}
\centering
\epsfig{file=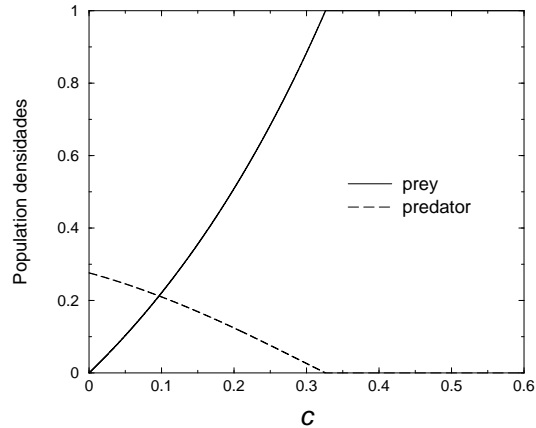,width=7cm}
\caption{Densities of predator and prey as functions
of $c$ for the quasi-spatial model, for $p=0.2$.}
\label{dens-two}	
\end{figure}

We have analyzed numerically 
the five-dimensional map, described by the set of equations
(\ref{pair1}), (\ref{pair2}), (\ref{pair3}), (\ref{pair4}) and (\ref{pair5}),
and we have obtained four types of solutions. Two solutions are
trivial and are given by $x=y=u=v=w=0$ and 
$x=1$, $y=u=v=w=0$. They correspond to the empty and 
prey absorbing states, respectively. The empty absorbing state, where 
both species have been extinct is an unstable solution and never occurs. 
However, the prey absorbing state is one of the possible stable 
stationary solutions and is stable above the critical transition line
$c=c_1(p)$ shown in figure \ref{diag-two}. Below this line
it becomes unstable giving rise to the active state.

The other solutions correspond to the active states where
both prey and predator coexist. These solutions are of two kinds: a
stationary solution where there is a coexistence of the two species with
densities constant in time, which we call the nonoscillating (NO) active
state; and another solution where both population densities oscillate in
time. This solution corresponds to a self-sustained oscillation of the
predator-prey system and will be called the oscillating (O) active state. 
In the phase diagram of figure \ref{diag-two} there is a line 
$c=c_2(p)$ that separates the NO and O active phases.
Figure \ref{dens-two} 
shows the behavior of the densities as a function of $c$ for $p=0$. 

\subsection{Oscillatory behavior}

\begin{figure}
\centering
\epsfig{file=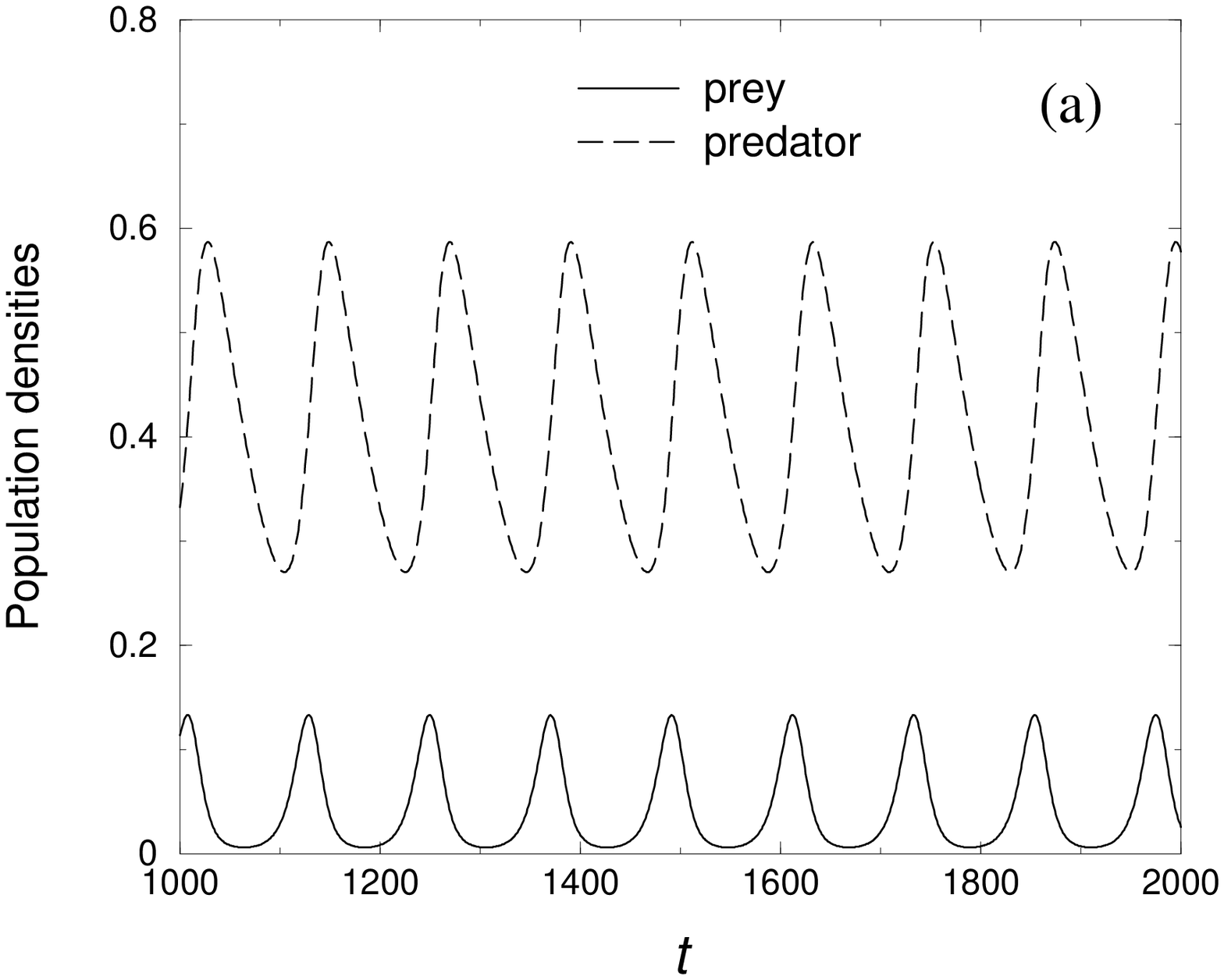,width=6.5cm}
\hfill
\epsfig{file=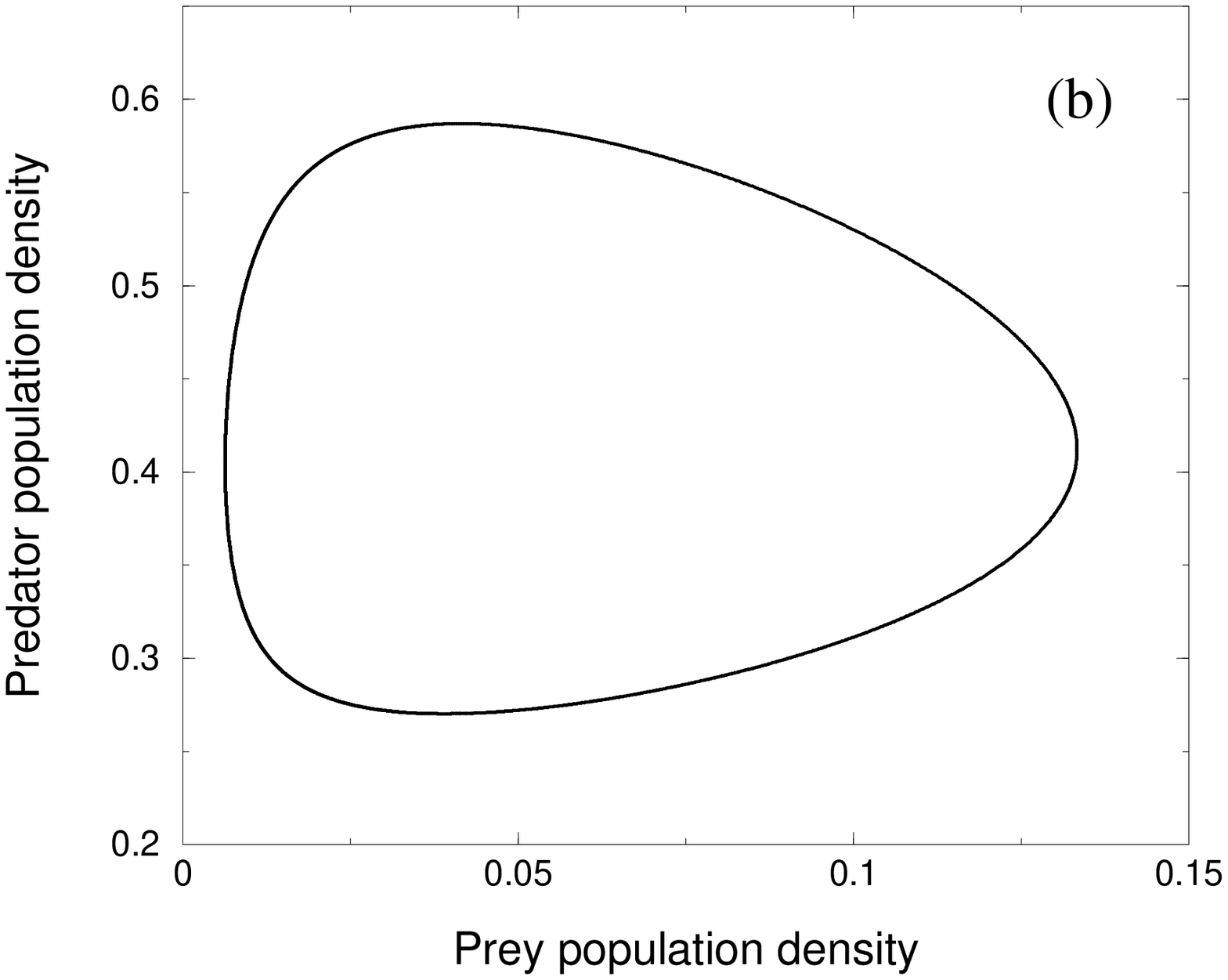,width=6.5cm}
\caption{(a) Densities of predator and prey as functions
of time and (b) density of predator versus density of prey,
for the quasi-spatial model, for $p=0$ and $c=0.016$.}
\label{osc-ab}
\end{figure}

In figure \ref{osc-ab} we show an example of self-sustained oscillations
of the densities of prey and predators as functions of time.
The oscillating solutions are attained from the nonoscillating solutions by a
Hopf bifurcation. The fixed point associated to this solution is an unstable
center which produces a stable limit cycle as trajectories in 
the phase-space of the predator density versus prey density, 
as can be seen in figure \ref{osc-ab}. Notice   
that the oscillations are not damped and have a
well defined period which is the same for the prey density and for the
predator density, which implies that the oscillations are coupled. A maximum
of predators always follow a maximum of prey. This means that the abundance
of prey is a condition that favors the increase in the number of
predators. As the predator number increases the prey population decays. The
evanescence of prey is followed by a decrease in the predator number,
giving conditions for the increase of prey population until the cycle 
starts again. 

A well defined oscillatory behavior is found for many biological
population, the most famous being the one related to the time
oscillations of the population of lynx and snowshoe hare in 
Canada for which data were collected for a long period of time 
\cite{hastings,ecology}. If the hare population cycles are mainly
governed by the lynx cycle then the oscillations shown by the present
model reproduces qualitatively some of the features of this 
predator-prey dynamics.

\begin{figure}
\centering
\epsfig{file=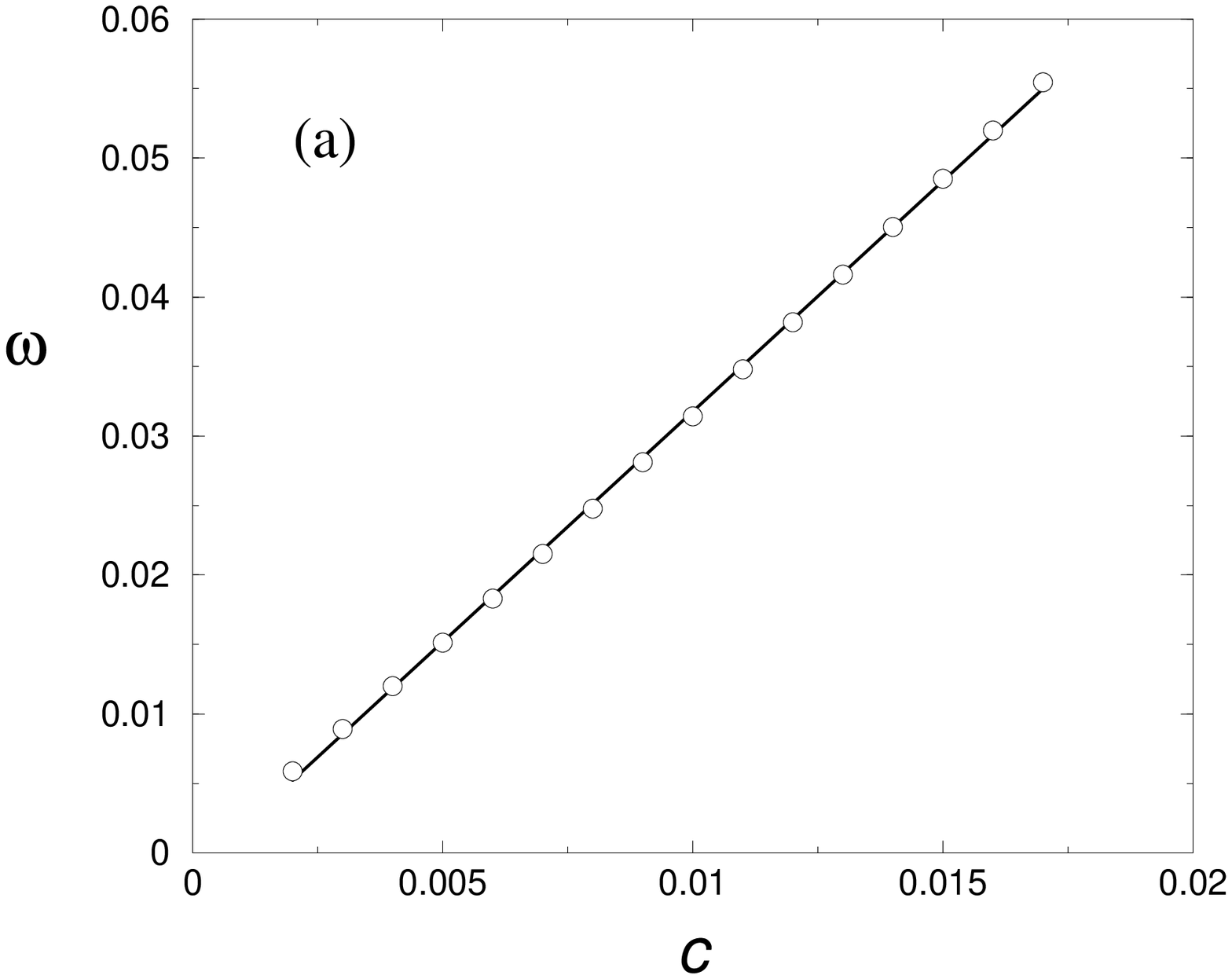,width=6.5cm}
\hfill
\epsfig{file=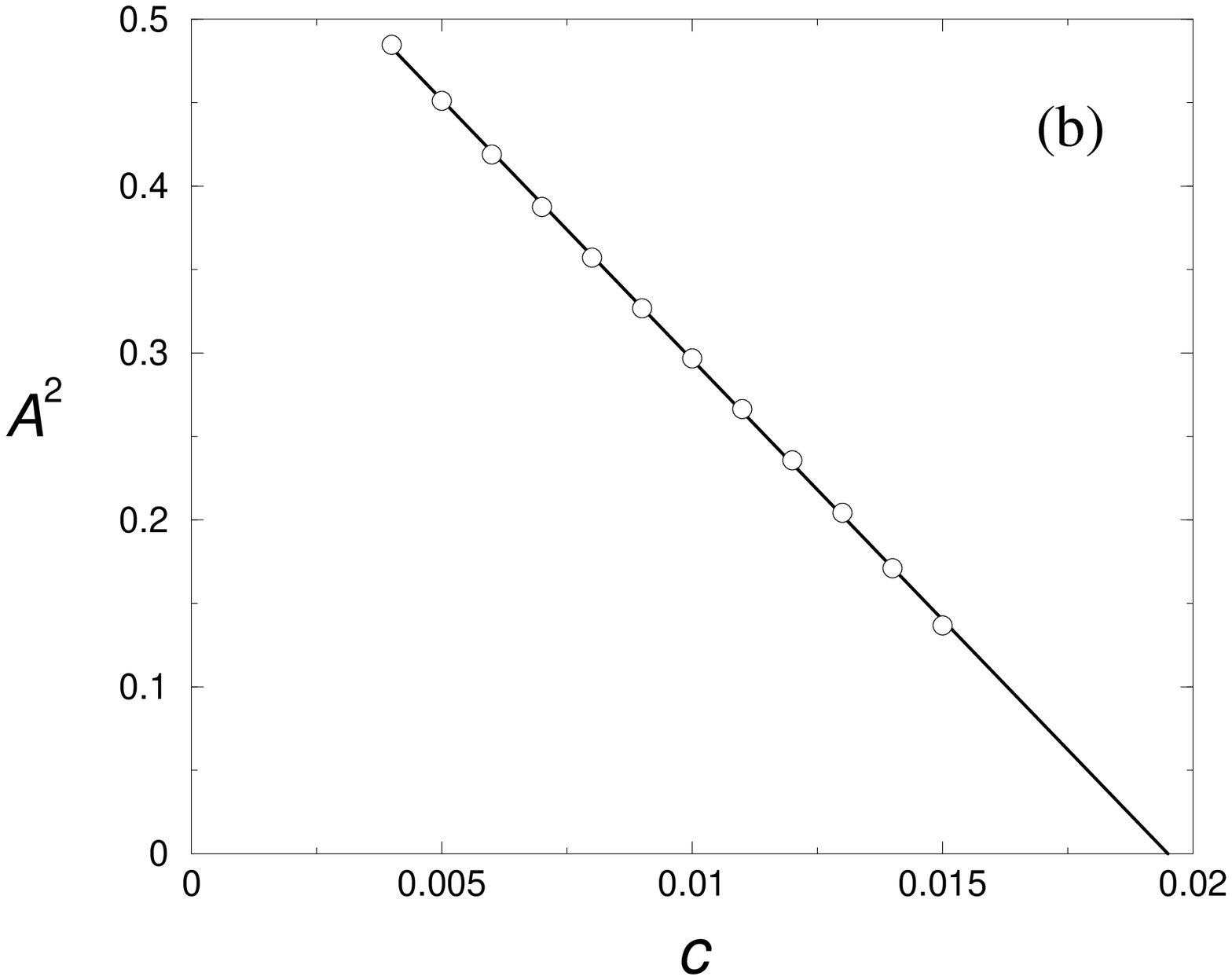,width=6.5cm}
\caption{(a) Frequency of oscillations $\omega$ versus the parameter $c$. 
The frequency vanishes linearly as one approaches $c=0$.
(b) Amplitude $A$ of oscillations versus $c$ near the Hopf bifurcation point
$c_2=0.019$. The quantity $A^2$ vanishes linearly when $c\to c_2$
in accordance with a Hopf bifurcation.}
\label{freq-ampl}
\end{figure}

Next we analyze the behavior of the frequency and amplitude
of oscillations.
Fixing the parameter $p$ and varying the parameter $c$, 
we verify that in all the
oscillating region the frequency of oscillation 
is proportional to parameter  $c$, 
\begin{equation}
\omega \sim c,  
\label{16}
\end{equation}
as can be seen in figure \ref{freq-ampl}. 
Low frequencies are associated to low values of 
$c$; what means that, for small values of $c$,
the greater the predator lifetime
the greater will be period of the oscillation. 
As to the amplitude $A$ of the oscillations, 
we have verified, that fixing the value of $p$ and
varying the parameter $c$, it increases as $c$ decreases. Our
results show that, 
\begin{equation}
A \sim (c-c_2)^{1/2},
\label{17}
\end{equation}
as expected for a Hopf bifurcation and shown in figure \ref{freq-ampl}.
The transition line $c=c_2$ from the oscillating phase to the nonoscillating
phase can either be obtained by using the criterion given by
equation (\ref{17}) or by analyzing the
eigenvalues associated to the map given by the set of equations
(\ref{pair1}), (\ref{pair2}), (\ref{pair3}), (\ref{pair4}) and 
(\ref{pair5}). This last criterion means to find the
points of phase diagram such that the real part of the
dominant complex eigenvalue equals $1$.

\section{Discussion and conclusion}

The main result coming from the pair mean-field approximation 
applied to the predator-prey PCA is that it is possible to describe 
coexistence and self-sustained time oscillations.
Moreover, these are stable oscillations. 
Given a set of parameters, just one limit
cycle is achieved, no matter what the initial conditions are. 
This property is essential in describing a biological system
since a small variation in the initial condition can not modify the 
amplitude, frequency and mean value of
the time oscillation densities of a predator-prey system. 
Similar results were obtained from a continuous
time version of the present model \cite{sat}.
Although the simple mean-field equations are essentially the 
same in both versions this is not the case concerning the 
pair mean-field approximation.
The time evolutions of the pair correlations for the PCA, 
presented here, depend on higher order correlations 
(up to fourth) when compared to the ones of the continuous version
(up to third). 

The model studied here is a spatial structured model with individuals
residing in sites of a lattice and described by discrete dynamic variables.
When we perform simple mean-field approximation we neglect all the
correlations of sites in the lattice. But we take into account that there
are limited resources for the surveillance of each species. For example in
the time evolution equation for the density of prey we have an explicit term
relative to reaction of birth of prey which is the product of the density of
prey $x$ by the density of empty sites $z=(1-x-y)$. This
coincides with an extended patch model approach for predator-prey systems.
The presence of this term is what differs the simple mean-field
equations from the Lotka-Volterra equations.
However, taking into account the limitation of space and resources
the simple mean-field equations are not sufficient to get
self-sustained oscillations although able to describe damped
time oscillations of population densities.

To get self-sustained time oscillations we had to proceed to
the next level of approximation in which a pair of nearest neighbor
sites is treated exactly. This approximation can be seen as
representing a pair of nearest neighbor sites immersed in a mean field
produced by the rest of the lattice. The most important feature
being the fact that the two sites of this pair can be seen as
localized in space. 
The set of five equations which results from the pair
approximation for the PCA is indeed able to produce self-sustained 
oscillations of population densities. It presents an important property
that the Lotka-Volterra model lacks, namely, the oscillating solutions 
are stable and are unique for a given set of the control parameters. 

\section*{Acknowledgements}

The authors have been supported by the Brazilian agency
CNPq.

\section*{References}


\end{document}